\documentstyle[12pt,a4wide]{article}

\title{\mbox{\LARGE{Primordial Magnetic Fields in False Vacuum Inflation}}}
\author{
\mbox{\large{Anne--Christine Davis}}
\thanks{e-mail: A.C.Davis@damtp.cam.ac.uk}
\vspace{0.2cm}\\
\mbox{\large{and}}
\vspace{0.2cm}\\
\mbox{\large{Konstantinos Dimopoulos}}
\thanks{e-mail: K.Dimopoulos@damtp.cam.ac.uk}
\vspace{0.4cm}\\
\mbox{\normalsize{\em Department of Applied Mathematics and
Theoretical Physics,}}\\
\mbox{\normalsize{\em University of Cambridge, Silver Street,}}\\
\mbox{\normalsize{\em Cambridge, CB3 9EW, U.K.}}
}

\begin{document}

\begin{titlepage}

\maketitle

\begin{abstract}
We show that, during false vacuum inflation, a primordial magnetic
field can be created, sufficiently strong to seed the galactic dynamo
and generate the observed galactic magnetic fields. Considering the
inflaton dominated regime, our field is produced by the Higgs--field
gradients, resulting from a grand unified phase transition. The
evolution of the field is followed from its creation through to the
epoch of structure formation, subject to the relevant constraints. We
find that it is possible to create a magnetic field of
sufficient magnitude, provided the phase transition occurs during the
final 5 \mbox{e-foldings} of the inflationary period. 
\end{abstract}

\vspace{2cm}

\begin{flushright}
CERN-TH/95-175\\
DAMTP-95-31\\
astro-ph/9506132
\end{flushright}

\end{titlepage}

\section{Introduction}

One of the most exciting astrophysical consequences of phase
transitions in the early universe is the possible creation of
primordial magnetic fields.

The existence of a primordial magnetic field could have significant
effect on various astrophysical processes. Indeed, large scale
magnetic fields are important in intercluster gas or rich clusters of
galaxies, in QSO's and active galactic nuclei. The existence of a
primordial field could influence the galaxy formation process and
would play a very important role in the resulting galactic spins
\cite{peeb}. A primordial field would also have an important effect
on the fragmentation process of large scale structure and of the
protogalaxies (by modifying the Jeans mass) and on the formation of
Population III stars \cite{hogan}. But the most important
consequence of the existence of a primordial magnetic field is that it
can seed the observed galactic magnetic field.

The galactic field is also very important to the astrophysics of the
galaxy. It influences the dynamics of the galaxy, the star formation
process (by transferring angular momentum away from protostellar
clouds \cite{zeld}, \cite{mestel} and by affecting the initial mass
function of the star formation process \cite{rees}), the dynamics of
compact stars (white dwarfs, neutron stars and black holes) and the
confining of cosmic rays, to name but some.

It is widely accepted that the galactic magnetic fields are generated
through a dynamo mechanism usually referred to as  {\em the galactic
dynamo}, for which, though, there is no consistent mathematical model
yet \cite{ref}. The basic idea of the dynamo mechanism is that a weak 
seed field could be amplified by the turbulent motion of ionised gas,
which follows the differential rotation of the galaxy \cite{zeld},
\cite{park}. The growth of the field is exponential and, thus, its
strength can be increased several orders of magnitude in only a few
e-foldings of amplification. 

The currently observed magnetic field of the Milky Way and of nearby
galaxies is of the order of a $\mu Gauss$. If the
e-folding time is no more than the galactic rotation period,
\mbox{$\sim 10^{8}yrs$} then, considering the galactic age,
\mbox{$\sim 10^{10}yrs$}, the seed field needed to produce  a field of the
observed value is about \mbox{$\sim 10^{-19}Gauss $} \cite{zeld},
\cite{rees} on a comoving scale of a protogalaxy.

Although, it has been argued by many authors that the seed field could
be produced from stars via stellar winds or supernovae and other
explosions \cite{zeld}, there is evidence that suggests that the seed
field is more likely to be truly primordial. For example, the observed
field of the Milky Way does not change sign with $z$ ($z$ being the
galactic altitude) as it would if it was produced by the stars of the
galactic disk \cite{zeld}. 

Various attempts have been made to produce a primordial field in the
early universe. A thorough investigation of the issue was
attempted by Turner and Widrow \cite{TW}, who incorporated inflation
and created the field by explicitly breaking the conformal invariance
of electromagnetism. This was done in a number of ways, such as 
coupling the photon to gravity through $RA^{2}$ and $RF^{2}$ terms
($R$ being the curvature, $A$ being the photon field and $F$ being the
electromagnetic field strength), or with a scalar field $\phi $, like
the axion, through a term of the form $\phi F^{2}$. It was, thus, shown
that satisfactory results could be obtained only at the expense of
gauge invariance. Garretson {\em et. al.} \cite{GFC} have generalized
the effort of \cite{TW} by coupling the photon to an arbitrary
pseudo-Goldstone boson, rather than the QCD axion. They have showed
however that, in all cases considered, it was impossible to generate a
primordial magnetic field of any astrophysical importance. Breaking the
electromagnetic conformal invariance during inflation was a mechanism
used also by a number of other authors, such as Ratra \cite{ratra} and
Dolgov \cite{dolg}. Ratra {\em has} been successful in generating an
adequately intense magnetic field. The field was generated by coupling
the field strength with a scalar field $\Phi $ (the dilaton) through a
term of the form $e^{\Phi}F^{2}$. Dolgov, however, did not introduce
any extra coupling but considered photon production by external
gravity by the quantum conformal anomaly. He
produced a field of enough strength, but  only in the case of a large
number (over 30) of light charged bosons.  

Another, more successful direction was using a phase transition for
the creation of a primordial field. An early effort was made by Hogan
\cite{hogan}, who considered the possibility of turbulence arising
during the QCD transition. His treatment, though, was based on a
number of assumptions concerning equipartition of energy, which are of
questionable validity. Much later Vachaspati \cite{Vach} proposed a
mechanism to produce a marginally sufficient magnetic field during the
electroweak transition.
This has also been addressed, a bit more successfully, by
Enqvist and Olesen \cite{Enqv}. The later have also considered a phase
transition to a new, ferromagnetic ground state of the vacuum, which
could also produce an adequately strong magnetic field \cite{EqOl}.
Finally, the literature contains a number of other, more exotic
mechanisms (such as, for example, the creation of a primordial
magnetic field by the turbulent motion of infalling matter into wakes
in the wiggly string scenario \cite{vilen}). In most of the cases,
though, the achieved field appeared to be too weak to seed the
galactic dynamo.

In this paper we examine the production of a primordial magnetic field
during false vacuum inflation. In false vacuum inflation a phase
transition can occur during the inflationary period. As shown by
Vachaspati \cite{Vach}, the existence of a horizon could result in the
production of a magnetic field at the phase transition. Although our
model incorporates the benefits of inflation, it does not require the
breaking of the gauge or even the conformal invariance of 
electromagnetism. Neither does it involve the addition of any extra
couplings between fields through the inclusion of peculiar, ``by
hand'' terms in the lagrangian. Our magnetic field is produced by the
dynamic features of the GUT Higgs--field, independently of specific
GUT models. No additional fields are introduced in the problem and the
results cover the most general case. By considering a GUT phase
transition we find that, for some parameter space, the magnetic field
produced is of enough strength to seed a galactic dynamo mechanism at
the epoch of structure and galaxy formation.  

Our results take into account
any constraints imposed on the field during
its evolution until the epoch of structure and galaxy formation.
Before examining the behaviour of the magnetic
field, we give a detailed description of the model and of the
mechanism, through which the original primordial magnetic field is
created. 

\section{Creation of the primordial magnetic field}

Vachaspati \cite{Vach} suggested that the existence
of a horizon would result in the creation of a primordial magnetic
field at a phase transition in the early universe.

Consider a non abelian group G. The field strength of the gauge fields
is,
\begin {equation}
H^{a}_{\mu\nu}=\partial_{\mu}X^{a}_{\nu}-
\partial_{\nu}X^{a}_{\mu}-g_{o}f^{a}_{bc}X^{b}_{\mu}X^{c}_{\nu}
\end{equation}
where $f^{a}_{bc}$ are the structure constants of G and $g_{o}$ is the
gauge coupling. If the symmetry of the gauge group G is broken,
leaving a residual symmetry corresponding to a subgroup H of G, then
the gauge fields of the residual symmetry are given by,
\begin{equation}
Y^{b}\equiv u^{ab}X^{a}
\end{equation}
where $u^{ab}$ is a unitary matrix specifying the directions of the
generators of the unbroken symmetry. Vachaspati argued that if the
symmetry was spontaneously broken then the VEV of the Higgs--field
$\psi^{a}$ would have been uncorrelated on superhorizon
scales\footnote{More precisely, on scales larger than the correlation
length.}, and hence, could not be ``aligned'' throughout all space
with a gauge transformation. Therefore, the gradients of the
Higgs--field would, in general, be non zero. Due to the coupling
through the covariant derivative,

\begin{equation}
D_{\mu}\langle\psi\rangle=
(\partial_{\mu}-ig\tau^{b}Y^{b}_{\mu})\langle\psi\rangle
\end{equation}
where $\tau^{a}$ are the generators of the residual symmetry and $g$
is the gauge coupling, the corresponding field strength $G^{b}_{\mu\nu}$
is non-zero. This can be seen explicitly by using the gauge invariant
generalization of 'tHooft \cite{Thof},
\begin{eqnarray}
G^{b}_{\mu\nu} & = &
u^{ab}[H^{a}_{\mu\nu}-g^{-1}\mu^{-2}f^{a}_{cd}D_{\mu}\psi^{c}D_{\nu}\psi^{d}]
\nonumber \\
& = &
\partial_{\mu}Y^{b}_{\nu}-\partial_{\nu}Y^{b}_{\mu}-g^{-1}
\mu^{-2}u^{ab}f^{a}_{cd}\partial_{\mu}\psi^{c}\partial_{\nu}\psi^{d}
\label{fstr}
\end{eqnarray}
where $\mu$ is the scale of the symmetry breaking. So
even if the gauge field $Y_{\mu}^{b}$ can be gauged away the field strength
is still non-zero:
\begin{equation}
G^{b}_{\mu\nu}=-\frac{1}{g\mu^2}
u^{ab}f^{a}_{cd}\partial_{\mu}\psi^{c}\partial_{\nu}\psi^{d}\label{gmn}
\end{equation}

Vachaspati applied the above in the case of the electroweak phase
transition, taking $G_{\mu\nu}$ to be the field strength of
electromagnetism.

In this paper we will follow a similar reasoning but for a GUT phase
transition. G is now the GUT symmetry group and the residual symmetry
is the electroweak. In order to get to electromagnetism we need to
consider also the final electroweak phase transition. In analogy with
the above the electromagnetic gauge potential is given by,
\begin{equation}
A_{\mu}=v^{b}Y^{b}_{\mu}\equiv
\sin\theta_{W} n^{a} W^{a}_{\mu}+\cos\theta_{W} B_{\mu}^{Y} \label{wein}
\end{equation}
where $v^{b} $ is a unit vector specifying the direction of the
unbroken symmetry $U(1)_{em} $ generator, $n^{a} $ are the $SU(2)$
generators of the electroweak group $SU(2)\times U(1)_{Y} $,
$W^{a}_{\mu}$ are the $SU(2)$ gauge fields, $B_{\mu}^{Y} $ is the
$U(1)_{Y} $ gauge field and $\theta_{W} $ is the Weinberg angle.
{}From~(\ref{wein}) it easy to see that,
\begin{equation}
v^{b}\equiv (\sin\theta_{W} n^{a} ,\cos\theta_{W} )
\end{equation}
\begin{equation}
Y^{b}_{\mu}\equiv (W^{a}_{\mu} ,B_{\mu}^{Y} ) \label{Yb}
\end{equation}
with $b=1,\ldots ,4$ and $a=1,\ldots ,3$

The contribution to the electromagnetic field stength $F_{\mu\nu} $ from
the GUT transition is, therefore, 

\begin{equation}
F_{\mu\nu}\equiv v^{b} G^{b}_{\mu\nu}\label{fmn}
\end{equation}

where $v^{b} G^{b}_{\mu\nu} $ is,
\begin{equation}
v^{b} G^{b}_{\mu\nu}\equiv\sin\theta_{W} n^{a} G^{a}_{\mu\nu}
+\cos\theta_{W} G^{Y}_{\mu\nu}
\end{equation}

with $b=1,\ldots ,4$, $a=1,\ldots ,3$ and $G^{Y}_{\mu\nu}\equiv
G^{4}_{\mu\nu} $.  

The magnetic field produced by the GUT phase transition is, therefore,
\begin{equation}
B_{\mu}\equiv\frac{1}{2}\varepsilon_{\mu\nu\lambda} F^{\nu\lambda}\label{bm}
\end{equation}

Thus, for an order of magnitude estimate, (\ref{gmn}), (\ref{fmn}) and
(\ref{bm}) suggest,
\begin{equation}
|B_{\mu}|\sim |F'_{\mu\nu} |\sim |G_{\mu\nu} |\sim\frac{1}{g\mu^{2} }
(\partial_{\mu}\langle\psi\rangle)^{2} \label{Btpsi}
\end{equation}
since $v^{a} $, $u^{ab} $ and
$f^{a}_{cd} $ are of unit magnitude. As far as the Higgs--field
gradients are concerned, on dimensional grounds, we have,
\begin{equation}
\partial_{\mu}\langle\psi\rangle\sim\frac{\mu}{\xi} \label{grad}
\end{equation}
where $\xi$ is the correlation length of the Higgs--field
configuration.

Also, at the GUT scale: \mbox{$4\pi g^{-2}\simeq 40 \Rightarrow
g^{-1}\sim 1$ }. Therefore, a dimensional estimate for the magnetic
field is,
\begin{equation}
B\equiv |B_{\mu} |\sim\xi^{-2}\label{mgfd}
\end{equation}

\section{False Vacuum Inflation}

In this section we review false vacuum inflation, a popular model of
inflation corresponding to extensive literature
(\cite{Lid1}...\cite{modl}). 

In this model the inflaton field $\phi$ rolls down its potential
towards the minimum, which does not correspond to the true vacuum, but
is instead a false vacuum state. There are two distinct and quite
different kinds of false vacuum inflation, depending on whether
the energy density is dominated by the false vacuum
energy density or by the potential energy density of the inflaton
field. Full details are given by Copeland {\em et al} in \cite{modl}.
Unlike them, we concentrate on the inflaton dominated case, as in
\cite{Kof1}...\cite{rest}. In this case, the phase transition does not
lead to the end of inflation as it does in the vacuum dominated case.

\subsection{The model}

In this model the energy density is dominated by a potential with two
scalar fields; the inflaton field $\phi$ and the Higgs--field $\psi$.
The later is responsible for the phase transition. We should emphasise
here that the Higgs--field considered does not correspond to a
specific GUT model and can have several components  without
this affecting the following analysis \cite{modl}.

We take the form of the potential to be,
\begin{equation}
V(\phi ,\psi )=\frac{1}{4}\lambda (\psi^{2} -\mu^{2} )^{2}
+\frac{1}{2} m^{2}\phi^{2} +\frac{1}{2}\lambda '\phi^{2}\psi^{2}\label{pot}
\end{equation}

The phase transition takes place at $\phi =\phi_{0} $, where,
\begin{equation}
\phi_{0}^{2}\equiv\frac{\lambda}{\lambda '}\mu^{2} \label{finst}
\end{equation}

This gives the effective scale of the symmetry breaking,
\begin{equation}
\mu^{2}_{e\!f\!f}\equiv\mu^{2} (1-\frac{\phi^{2} }{\phi^{2}_{0} }) \label{yVEV}
\end{equation}

Without loss of generality  we  assume that $\phi $ is
initially positive and rolls down the
potential in such a way that \mbox{$\dot{\phi} < 0 $}. 
If there is sufficient inflation before the phase transition and
$\lambda\gg\lambda'$ the Higgs--field will have rolled to the minimum
of its potential $\psi=0$ before the inflaton falls to its critical
value $\phi_{0}$. So, when $\phi >\phi_{0} $,
\begin{equation}
V(\phi ,0)=\frac{1}{4}\lambda\mu^{4} +\frac{1}{2} m^{2} \phi^{2} \label{fpot}
\end{equation}

In the slow roll approximation 
the dynamics of inflation are governed by the equations,

\begin{eqnarray}
H^{2} & \simeq & \frac{8\pi }{3m^{2}_{pl} } V \label{Hsqr}\\
3H\dot{\phi } & \simeq & -V' \label{fdot}
\end{eqnarray}
where the prime and the dot denote derivatives with respect to $\phi $
and time respectively,  
\mbox{$H\equiv\dot{a} /a $} is the Hubble parameter, $a$ is the
scale factor of the Universe, and 
$m_{pl}$ is the Planck mass (\mbox{$m_{pl}=1.22\times 10^{19}GeV$}).

Thus,
\begin{equation}
H \simeq - \frac{8\pi }{m^{2}_{pl} } \frac{V}{V'} \dot{\phi } \label{HVV}
\end{equation}

 From (\ref{HVV}) the number of
e-foldings of expansion, which occur between the values $\phi_{1} $ and
$\phi_{2} $ of the inflaton field, is given by,
\begin{equation}
N(\phi_{1} ,\phi_{2} ) \equiv \ln\frac{a_{2} }{a_{1} } \simeq
-\frac{8\pi }{m^{2}_{pl} } \int^{\phi_{2} }_{\phi_{1} } \frac{V}{V'}
d\phi \label{Nf1f2}
\end{equation}

\subsection{The inflaton dominated regime}

This is the case we are going to be interested in, since
{\em the back reaction of the Higgs--field $\psi $ to the inflaton
field $\phi $ is negligible} and so {\em the phase transition does not
cause the end of inflation} \cite{modl}. If the opposite is true 
and inflation ends at the phase transition, then the effects of the
transition are not too different from the usual, thermal phase
transitions studied in the literature.

In the inflaton dominated case the energy density of the inflaton
field in (\ref{fpot}) is much larger than the false vacuum energy
density. Therefore,
\begin{equation}
V(\phi )\simeq \frac{1}{2} m^{2} \phi^{2}\label{chao}
\end{equation}
which is identical with chaotic inflation. 
Inflation ends at $\phi \sim \phi_{\varepsilon}$, where,

\begin{equation}
\phi_{\varepsilon}\equiv\frac{m_{pl}}{\sqrt{4\pi}}\label{fend}
\end{equation}

To ensure inflaton domination until the end of inflation then,
\begin{equation}
\frac{1}{2} m^{2} \phi^{2}_{\varepsilon} \gg \frac{1}{4}\lambda \mu^{4}
\Rightarrow \frac{2\pi }{m^{2}_{pl} } \frac{\lambda\mu^{4} }{m^{2} }
\ll 1 \label{infdom}
\end{equation}
This is the condition for inflaton domination. At this point it
should be mentioned that {\em if the inflaton domination condition is
strongly valid
the dynamics of inflation are not seriously affected by the phase
transition, provided that $\psi $ falls rapidly to its VEV}.
Thus, the equations (\ref{Hsqr}), (\ref{fdot}) and
(\ref{HVV}) can be used throughout the duration of inflation.

Thus, for the number of e-foldings between the phase transition and
the end of inflation, in the inflaton dominated case,  we obtain,
\begin{equation}
N\equiv N(\phi_{0} ,0 )=\frac{2\pi }{m_{pl}^{2} } \phi_{0}^{2} \label{No}
\end{equation}
where we have used $\phi_{0} \gg \phi_{\varepsilon}$.

Finally, for the roll down of the inflaton field, using (\ref{Hsqr}),
(\ref{fdot}) and (\ref{chao}) we obtain,
\begin{equation}
\dot{\phi } = -\frac{m_{pl}\, m}{\sqrt{12\pi } }\label{roll}
\end{equation}

\section{The correlation length}

\subsection{Evolution of the correlation length}

Through the use of the uncertainty principle, we can estimate the range
of any interaction. Therefore, the physical correlation length 
for the Higgs--field $\psi $ is,
\begin{equation}
\xi \equiv \frac{1}{|m_{H}|}
\label{ximH}
\end{equation}
where $m_{H} = m_{H}(t) $ is the mass of the Higgs particle,
\begin{equation}
m_{H}^{2}=\frac{\partial^{2}V(\phi ,\psi )}{\partial\psi^{2} }=
3\lambda\psi^{2}+\lambda'\phi^{2}-\lambda\mu^{2}
\label{mH}
\end{equation}
Before the phase transition $\psi =0 $ and thus, \mbox{$m_{H}^{2} =
\lambda|\mu_{e\!f\!f}^{2}|$}. Therefore, the physical correlation
length is, 
\begin{equation}
\xi (\phi ) = \frac{1}{\sqrt{\lambda }\,\mu }\,\frac{\phi_{0} }{\sqrt{\phi^{2}
-\phi_{0}^{2} } } \label{xitf}
\end{equation}

However, this is not valid as
we approach the phase transition, \mbox{$\phi \rightarrow \phi_{0}$}.
As \mbox{$\dot{\xi } =1$} the correlation length grows linearly with
time as shown in Figure 1.

Define,
\begin{equation}
\tau \equiv t-t_{0} \label{tau}
\end{equation}
where $t_{0}$ is the time the transition occurs.

If at $\tau\equiv\tau_{H} $,
\begin{equation}
\frac{d}{d\tau }(\frac{1}{|m_{H}|})
\,\vline_{_{_{_{_{_{\,H} } } } } } =1 
\label{caus}
\end{equation}

Then, from $\tau_{H}$ until the transition $\tau = 0$,
\mbox{$\frac{d\xi }{d\tau } =1$} and, therefore, the correlation
length $\xi_{0} $ at the time of the transition is, 
\begin{equation}
\xi_{0} = \xi_{H} -\tau_{H} \label{corr}
\end{equation}
where $\xi_{H} $ is the correlation length at $\tau_{H} $, and in the
linear regime, 
\begin{equation}
\xi(\tau)=\xi_{0}+\tau
\label{xilin}
\end{equation}

The linear growth of $\xi$ continues until it hits the declining slope
of \mbox{$|m_{H}|^{-1}$} (Figure 1). From then on the correlation length is given
again by (\ref{ximH}).

The phase transition we are considering is not triggered by
temperature fall, but by the roll down of the inflaton field. In that
sense {\em it is not a thermal phase transition}. Also, since it
occurs during inflation, the Universe is in a supercooled state
with temperature $T\approx 0$ and so {\em there are no Ginzburg
phenomena}. However, the configuration of the Higgs--field $\psi $
does not freeze at the moment of the phase transition due to long wave
quantum fluctuations that dominate the Higgs--field evolution
immediately after the transition.

The long wave fluctuations of $\psi$ are determined by the behaviour
of the Higgs--field mass (\ref{mH}). Immediately after the phase
transition the Higgs--field is still \mbox{$\psi\simeq 0$}
\cite{pogos}. Then, since \mbox{$\phi<\phi_{0}$}, it follows from
(\ref{mH}) that \mbox{$m_{H}^{2}<0$}, and quantum fluctuations grow
until the evolution of $\psi$ becomes potential
dominated and the field starts falling to its new minima. At this
stage the fluctuations of the Higgs--field become impotent and the
field configuration topology freezes. The fall of $\psi$ is very rapid
\cite{pogos}. After, \mbox{$\psi=\mu_{e\!f\!f}/\sqrt{3}$}, the mass of
the field becomes positive again (\ref{mH}). When the field reaches
its minimum \mbox{$\psi=\mu_{e\!f\!f}$}, then,
\mbox{$m_{H}^{2}=2\lambda\mu_{e\!f\!f}$}. 

The magnetic field is formed at the freezing of the Higgs--field
configuration. This occurs when \mbox{$|m_{H}^{2}|\simeq H^{2}$},
\cite{Kof1}, \cite{hodge}.\footnote{Nagasawa and Yokoyama \cite{yoko}
suggest that the freezing of the field occurs a bit later. However,
with the set of parameters used (see 8.1), the corresponding
difference in the correlation length is less than an order of
magnitude.} The correlation length $\xi_{F}$ at the 
time of freezing $\tau_{F}$, is either given by (\ref{ximH}) or by
(\ref{xilin}), depending on whether we are still in the linear regime
of not. 

\subsection{Computation of $\xi_{0}$ and $\xi_{F}$}

We assume that $\phi_{H} $ is very close to
$\phi_{0} $, or, equivalently, that the time $\tau_{H} $, when the
growth of $\xi $ reaches the speed of light, is very close to the time
$\tau =0$ of the phase transition. Therefore,
\begin{equation}
\Lambda\equiv\frac{\phi_{H}^{2} -\phi_{0}^{2} }{\phi_{0}^{2} } \ll 1
\label{asum}
\end{equation}

This will be verified when we introduce specific values for the parameters.

Using (\ref{roll}) and (\ref{xitf}) we find,
\begin{equation}
\frac{d}{d\tau }(\frac{1}{|m_{H}|})\,\vline_{_{_{_{_{_{\,H} } } } } } =
\frac{m_{pl}\, m}{\sqrt{12\pi } }\,\frac{1}{\sqrt{\lambda'}
}\,\frac{\phi_{0} }{[\phi_{H}^{2} -\phi_{0}^{2}]^{3/2} }
\end{equation}

Using, (\ref{caus}) we obtain,
\begin{equation}
\phi_{H}^{2} -\phi_{0}^{2} = [\frac{m_{pl}\, m\, \phi_{0} }{\sqrt{12\pi
\lambda'} }]^{2/3} \label{brac}
\end{equation}

 From (\ref{roll}) we have,
\begin{equation}
\phi_{H} = \phi_{0} - \frac{m_{pl}\, m}{\sqrt{12\pi } }\,\tau_{H}
\label{evoltau}
\end{equation}

Thus, solving for $\tau_{H} $ we obtain,
\begin{equation}
\tau_{H} = -\sqrt{3\pi } (12\pi \lambda' m_{pl}\, m\, \phi_{0} )^{-1/3}
\label{tauH}
\end{equation}

and thus,
\begin{equation}
\xi_{H} = \frac{1}{\sqrt{\lambda'} }\,[\frac{m_{pl}\, m\, \phi_{0}
}{\sqrt{12\pi \lambda' } }]^{-1/3} = 2\sqrt{3\pi }(12\pi\lambda'
m_{pl}\, m\, \phi_{0} )^{-1/3} \label{xiH}
\end{equation}
Using (\ref{corr}) and the above we obtain,
\begin{equation}
\xi_{0} = 3\sqrt{3\pi } (12\pi\lambda' m_{pl}\, m\, \phi_{0} )^{-1/3}
\label{corl}
\end{equation}

Let us compute, now, the correlation length at the time when the field
configuration freezes, i.e. when \mbox{$|m_{H}^{2}|\simeq H^{2}$}.
From (\ref{Hsqr}), (\ref{chao}) and (\ref{mH}) we obtain,

\begin{equation}
\phi_{0}^{2}-\phi_{F}^{2}\simeq \frac{4\pi
}{3}\,\frac{m^{2}\phi_{0}^{2}}{\lambda'm_{pl}^{2} }
\label{f0ff}
\end{equation}
where $\phi_{F}$ is the magnitude of the inflaton at the time of
freezing $\tau_{F}$, for which we find,

\begin{equation}
\tau_{F}\simeq\sqrt{3\pi
}\,\frac{(\phi_{0}^{2}-\phi_{F}^{2})}{m_{pl}m\phi_{0}}\simeq
\sqrt{\frac{16\pi^{3}}{3}}\,\frac{m\phi_{0}}{\lambda'm_{pl}^{3}}
\label{tf}
\end{equation}
where we have used (\ref{roll}) and the assumption (\ref{asum}).

Using (\ref{Nf1f2}), (\ref{chao}) and (\ref{f0ff}) we find that the
number of e-foldings of inflation between the phase transition and the
freezing of the field configuration is given by,

\begin{equation}
\Delta
N_{F}\simeq\frac{8\pi^{2}}{3}\frac{m^{2}\phi_{0}^{2}}{\lambda'm_{pl}^{4}} 
\label{DN}
\end{equation}

Only during this time are the quantum fluctuations of the Higgs--field
important.\footnote{During this time the quantum fluctuations of the
Higgs--field could create a ``mountain'' on the spectrum of the adiabatic
density perturbations from the inflaton field (as well as isocurvature
fluctuations from the Higgs--field) \cite{Kof1}, \cite{pogos}. The
scale of the perturbations is determined by the Horizon size at the
time of the transition. If the transition occurs when the
observable scales leave the horizon (about 60-45 e-foldings before the
end of inflation), there is a chance of producing excessive density
fluctuations and microwave background anisotropies. For a detailed
review see Salopek {\em et al.} \cite{rest}.} If \mbox{$\Delta
N_{F}\leq 1$} then the transition proceeds rapidly \cite{hodge}.

If we are still in the regime of linear growth of the correlation
length the value of it at the time of freezing is simply given by
(\ref{xilin}),
\begin{equation}
\xi_{F}^{(1)}=\xi_{0}+\tau_{F}
\label{xifl}
\end{equation}

If, however, the linear growth of $\xi$ has ended before $\tau_{F}$,
then it is given by (\ref{ximH}),

\begin{equation}
\xi_{F}^{(2)}\simeq \sqrt{\frac{3}{8\pi}}\,\frac{m_{pl}}{m\phi_{0}}
\label{xifm}
\end{equation}
where we have used that, \mbox{$m_{H}^{2}\simeq\lambda'(
\phi_{0}^{2}-\phi_{F}^{2})$} and (\ref{f0ff}).

Therefore, the initial correlation length of the Higgs--field
configuration is given by,

\begin{equation}
\xi_{F}=\mbox{min}(\xi_{F}^{(1)},\xi_{F}^{(2)})
\label{xF}
\end{equation}

\input epsf

\begin{figure}
\begin{center}
\leavevmode
\hbox{%
\epsfxsize=6.0in
\epsffile{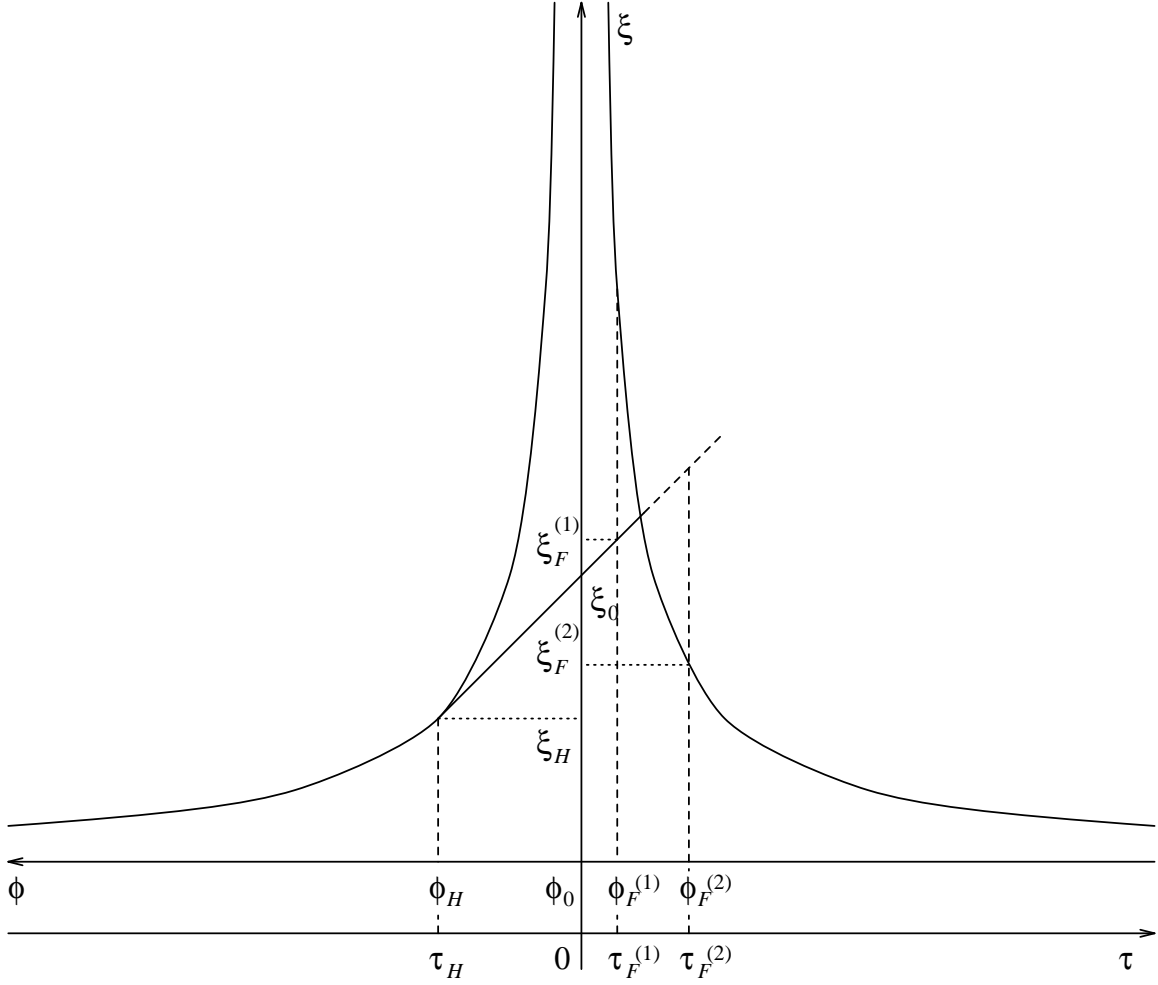}}
\end{center}
\caption{Evolution of the correlation length $\xi$ of the Higgs--field
configuration near the phase transition. The linear growth of $\xi$
starts at $\tau_{H}$. $\xi_{0}$ is the correlation length at the time
of the transition, $\xi_{F}^{(1)}$ is the correlation length at the
time the Higgs--field configuration freezes, when this occurs during
linear growth, and $\xi_{F}^{(2)}$ is the correlation length at freezing
when this occurs after the end of the linear regime.}
\end{figure}

\section{Evolution of the magnetic field}

\subsection{During inflation}

 From (\ref{xF}) and (\ref{mgfd}) we estimate the initial value of the
magnetic field produced just after the phase transition to be,
\begin{equation}
B_{0}\sim \xi_{F}^{-2}
\label{B0}
\end{equation}

Of course, after the GUT phase transition we still have
electroweak unification. Therefore (\ref{B0}) represents, in fact,
an ``electroweak'' magnetic field. However,
since the residual, electromagnetic symmetry generator is just a
projection of the electroweak generators (through the Weinberg angle),
the residual ``electromagnetic'' magnetic field will be of the same order
of magnitude as the one given by (\ref{B0}). Therefore, from now on, we
will ignore the electroweak transition and treat the above magnetic
field as ``electromagnetic''.

During inflation the magnetic field is {\em not} frozen into the
supercooled plasma \cite{TW} but still it scales as $a^{-2} $, since
it remains coupled to the Higgs--field $ \psi $ and, thus,
(\ref{mgfd}) is still valid. The configuration of the Higgs--field
remains comovingly frozen during inflation because the initially
correlated volumes expand exponentially, faster than causal
correlations. This is not the case after inflation ends. The scale
factor, then, grows slower than the causal correlations and the
comoving picture of the Higgs--field configuration starts changing
as the field becomes correlated over larger and larger comoving
volumes. However, after the end of inflation the magnetic field gets
frozen into the reheated plasma\footnote{see also section 7.5}
and decouples from the Higgs--field. 

\subsection{ The rms magnetic field}

In order to estimate the magnetic field on scales larger than the
typical dimensions of the correlated volumes, we have to introduce a
statistical method to do so. 

A thorough treatment by Enqvist and Olesen \cite{Enqv} suggests
that, in all cases that the Higgs--field gradients are a diminishing
function of $n$ (number of correlated domains), the root mean square
value of the field would behave as,
\begin{equation}
B_{rms} \equiv \sqrt{\langle B^{2} \rangle } = \frac{1}{\sqrt{n} }B_{cd}
\label{Brms}
\end{equation}
where $B_{cd}$ is the field inside a correlated domain and $n$ is the
number of correlation lengthscales, over which the field is averaged.
In their treatment Enqvist and Olesen choose the Higgs-field gradients
as the stochastic variables and also assume that their distribution is
gaussian and isotropic. Choosing the magnetic field itself as the
stochastic variable, Enqvist and Olesen reached the same result
(\ref{Brms}). 

At this point it should be mentioned that in the above treatment the rms
value of the field has been computed as a line average, that is an
average over all the possible curves in space between the points that
fix the lengthscale, over which the field is averaged. The above
result may be sensitive to the averaging procedure. One argument in
favour of line-averaging is that the current galactic magnetic field
has been measured using the Faraday rotation of light spectra, which
is also a line (line of sight) computation. If we assume that the
ratio of the seed field for the galactic dynamo and the currently
observed galactic field is independent of the averaging procedure then
this would suggest that line-averaging is required for the computation
of the primordial field. However, the nonlinearity of the dynamo
process as well as the rather poor knowledge we have for galaxy
formation make such an assumption non-trivial. In any case,
apart from the above, there seem to be no other argument in favour of
a particular averaging procedure. Therefore, using line-averaging
could be the safest choice. 
Here it is important to point out that
{\em line averaging just gives an estimate of the rms field and does
not correspond to any physical process}.

Suppose that we are interested in calculating the rms field at a time
$t$ over a physical lengthscale $L=L(t)$. Then,
\begin{equation}
n(t)=\frac{L}{\xi } \label{nLxi}
\end{equation}
where $\xi $ is the correlation length. This scale is equal to
the correlation length $\xi_{0} $ at the time of the phase transition.

In this paper we are mainly interested in the value of the magnetic field at
$t_{eq}$, the time of equal matter and radiation densities,
when structure formation begins. The scale of interest is the typical
intergalactic distance, since $t_{eq}$ is preceeding the gravitational
collapse of the galaxies (see also 7.2). At $t_{eq}$ the corresponding
scale is found to be,  

\begin{equation}
L_{eq}\sim (\frac{t_{eq} }{t_{p}})^{^{2/3} }L_{p}\sim 10\,pc \label{Leq}
\end{equation}
where $t_{p}\sim 10^{18}sec$ is the present time and
\mbox{$L_{p}\sim 1\,M\!pc$} is taken as the typical intergalactic
scale at present. 

From (\ref{nLxi}) the number of correlated domains at $t_{eq}$ is,
\begin{equation}
n\equiv n_{eq}=\frac{L_{eq} }{\xi_{eq} } \label{n}
\end{equation}
where $\xi_{eq}$ is the correlation length at $t_{eq}$.

Therefore, the rms value of the magnetic field over the scale of a
protogalaxy is:
\begin{equation}
B^{eq}_{rms}\sim \frac{1}{\sqrt{n} }B^{eq}_{cd} \label{brms}
\end{equation}
where $B^{eq}_{cd}$ is the value of the field inside a correlated domain at
$t_{eq}$.

\subsection{Growth of the correlated domains}

It is clear that, in order to calculate the rms field over the galactic
scale at $t_{eq}$ we need to estimate the correlation length $\xi_{eq}$, 
i.e. the size of the correlated domains at that time. Therefore, we
have to follow carefully the evolution of the correlated domains
throughout the whole radiation era.\footnote{The correlated domains
should not be pictured as attached bubbles of coherent magnetic field,
but as regions arround any given point in space in which the
orientation of the field is influenced by its orientation at this
point.}

During inflation, as explained already, $\xi $ scales as the scale factor
$a$. However, after the end of inflation, it grows faster. 
This is because, when two initially uncorrelated neighbouring domains
come into causal contact, the magnetic field around the interface is
expected to untangle and smooth, in order to avoid the creation of 
energetically unfavoured magnetic domain walls. In time the field
inside both domains ``aligns'' itself and becomes coherent over the
total volume. The velocity $v$, with which such a reorientation
occurs, is determined by the plasma, which carries the field and has
to reorientate its motion for that purpose.\footnote{Note that the
plasma does not have to be carried from one domain to another or get
somehow mixed. Also, conservation of flux is not violated with the
field's rearrangements, since {\em the field always remains frozen
into the plasma}, which is carried along.} 

Thus, the evolution of the correlation length is given by,

\begin{equation}
\frac{\mbox{d}\xi }{\mbox{d}t}=H\xi +v \label{dxi}
\end{equation}
where $v$ is the peculiar,
bulk velocity, determined, in principle, by the state of the plasma. 
 
From (\ref{dxi}) it is apparent that the correlated domains could grow
faster than the Hubble expansion. Therefore, {\em the magnetic field
configuration is not necessarily comovingly frozen} and the domains
could expand much faster than the universe, resulting in large
correlations of the field and high coherency. 

In order to describe the evolution of the correlated domains one has
to determine the peculiar velocity $v$. This primarily depends on
the opacity of the plasma. 

If the plasma is opaque on the scale of a correlated domain, then
radiation cannot penetrate this scale and is blocked inside the plasma
volume. Consequently, the plasma is subject to the total magnetic
pressure of the magnetic field gradient energy. Therefore, this energy
dissipates through coherent magnetohydrodynamic oscillations, i.e.
Alfven waves. In this case, the peculiar velocity of the magnetic field
reorientation, is the well known
Alfven velocity \cite {hogan},\footnote{Unless explicitly specified,
natural units are being used (\mbox{$\hbar =c=1$}). In natural units, \mbox{$G=m_{pl}^{-2}$}.}

\begin{equation}
v_{A}\equiv
\frac{B_{cd}}{\sqrt{\rho }} \label{valf}
\end{equation}
where $B_{cd}$ is the magnitude of the magnetic field inside a
correlated domain and $\rho $ is the {\em total} energy density of
the universe, since,
before $t_{eq}$, matter and radiation are strongly
coupled.\footnote{This coupling implies that any reorientation of the
momentum of matter has to drag radiation along with it. This
increases the inertia of the plasma, that balances the magnetic
pressure.}

If the plasma is not opaque over the scale $\xi $ of a correlated
domain, then radiation can penetrate this scale and carry away
momentum, extracted from the plasma through Thomson scattering of the
photons. This subtraction of momentum is equivalent to an effective
drag force, \mbox{$F\sim\rho\,\sigma_{T}\,v_{T}\,n_{e}$} \cite{hogan}.
Balancing this force with the magnetic force determines the
``Thomson'' velocity over the scale $\xi $,

\begin{equation}
v_{T}\equiv\frac{v_{A}^{2} }{\xi n_{e}\sigma_{T} } \label{vt}
\end{equation}
where $v_{A}$ is the Alfven velocity, $n_{e}$ is the electron number
density and $\sigma_{T}$ is the Thomson cross-section.

Hence, for a non--opaque plasma the peculiar velocity
of the plasma reorientation is given by \cite{hogan},

\begin{equation}
v=\mbox{min}(v_{A},v_{T})
\label{v}
\end{equation}

In order to explore the behaviour of the opaqueness of the plasma, we
need to compare the mean free path of the photon \mbox{$l_{T}\sim
(n_{e}\sigma_{T})^{-1}$} to the scale $\xi $ of the correlated
domains. For realistic models, the correlated
domains remain opaque at least until the epoch \mbox{$t_{anh}\sim
0.1\,sec$} of electron--positron annihilation \mbox{($T\sim
1\,M\!eV$)}. The reason for this can be easily understood by
calculating $l_{T}$ before and after pair annihilation.

For \mbox{$T>1\,M\!eV$}, instead of the usual Thomson cross-section
$\sigma_{T}$, we  have the Klein-Nishina cross-section
\cite{KN},  

\begin{equation}
\sigma_{KN}\simeq \frac{3}{8}\,\sigma_{T}\,(\frac{m_{e}}{T})\,[\,\ln
\frac{2T}{m_{e}}+\frac{1}{2}\,]
\simeq 2.7\,(\frac{GeV}{T})\,\ln [\frac{T}{GeV}]\,GeV^{-2}
\label{sKN}
\end{equation}
where \mbox{$m_{e}\simeq 0.5\,GeV$}is the electron mass and
\mbox{$\sigma_{T}\simeq 6.65\times 10^{-25}cm^{2}\simeq
1707.8\,GeV^{-2}$}. The electron number density is given by \cite{KT},

\begin{equation}
n_{e}\simeq \frac{3}{4}\frac{\zeta(3)}{\pi^{2}}\,g_{e}T^{3}
\label{ne}
\end{equation}
where \mbox{$\zeta(3)\simeq 1.20206$} and
\mbox{$g_{e}=4$} are the internal degrees of freedom of
electrons and positrons.

From (\ref{sKN}) and (\ref{ne}) we find,

\begin{equation}
l_{T}\sim\frac{0.1\,GeV}{T^{2}}\;\;\;\;\;\;\;\;\;\;\mbox{for
$T>1\,M\!eV$}
\end{equation}
which at annihilation gives, \mbox{$l_{T}(t_{anh})\sim
10^{5}GeV^{-1}$}.

After annihilation the electron number density is given by
\cite{KT},

\begin{equation}
n_{e}\simeq 6\times 10^{-10}n_{\gamma }\simeq 1.44\times 10^{-10}T^{3}
\label{NE}
\end{equation}
where $n_{\gamma }$ is the photon number density given by,

\begin{equation}
n_{\gamma }\simeq \frac{\zeta(3)}{\pi^{2}}\,g_{\gamma }T^{3}
\label{ng}
\end{equation}
where \mbox{$g_{\gamma }=2$} are internal degrees of freedom of the
photon. 

With the usual value for $\sigma_{T}$ we obtain,
\begin{equation}
l_{T}\sim\frac{10^{6}GeV^{2}}{T^{3}}\;\;\;\;\;\;\;\;\;\;\mbox{for
$T<1\,M\!eV$}
\end{equation}

At annihilation the above gives, \mbox{$l_{T}(t_{anh})\sim
10^{15}GeV^{-1}$}. 

Hence, the mean free path of the photon at the time of pair
annihilation is enlarged by a factor of $10^{10}$! As a result,
$l_{T}$ is very likely to become larger than $\xi$ after $t_{anh}$.
If this is so, the Thomson dragging effect has to be taken into
account and the peculiar velocity of the plasma reorientation is given
by (\ref{v}).

In order to calculate the peculiar velocity it is necessary to compute
the Alfven velocity, which requires the knowledge of the magnetic
field value $B_{cd}$ inside a correlated domain. To estimate this 
we assume that the magnetic flux, on scales larger than
the sizes of the correlated domains, is conserved, as implied by the
frozen--in condition.

Consider a closed curve $C$ in space, of lengthscale \mbox{$L>\xi$},
encircling an area $A$. Conservation of flux suggests
that the flux averaged mean magnetic field inside $A$ scales as
$a^{-2}$.  This implies that for the
field inside a correlated domain we have,
\mbox{$B_{cd}(L/\xi)^{-1}\propto a^{-2}$}.\vline\footnote{Note that
the flux averaging of the field on scales larger than the
correlation length, {\em corresponds to a physical process}, that of
the field untangling, and is so in order to preserve flux conservation
on scales that the magnetic field is frozen into the plasma. This
should not be confused with the line averaging procedure which we use to
estimate the rms field, and does {\em not} correspond to a physical
process.}
Since $C$ follows the
universe expansion \mbox{$L\propto a$}, with \mbox{$a\propto t^{1/2}$}.
Thus, for the radiation era, we obtain,

\begin{equation}
B_{cd}\,t^{1/2}\xi=K\Rightarrow B_{cd}=\frac{K}{t^{1/2}\xi}
\label{K}
\end{equation}
where $K$ is a constant to be evaluated at any convenient time.
Since, the correlation length grows at least as fast as the
universe expands, the magnetic field inside a
correlated domain dilutes at least as rapidly as $a^{-2}$ for the
radiation era.

Substituting the above into (\ref{valf}) we find,

\begin{equation}
v_{A}\sim 10\,\frac{K}{m_{pl}}\,\frac{t^{1/2}}{\xi}
\label{vA}
\end{equation}

Solving the evolution equation (\ref{dxi}) with \mbox{$a\propto
t^{1/2}$} in the case that \mbox{$v=v_{A}$} gives,

\begin{equation}
\xi(t)^{2}=(\frac{t}{t_{i}})\,\xi_{i}^{2}+4v_{A}(t)\,\xi(t)\,
t\,(1-\sqrt{\frac{t_{i}}{t}}\,) 
\label{xivA}
\end{equation}
where $\xi_{i}$ is the correlation length of the field at the time
$t_{i}$. The first term of (\ref{xivA}) is due to the Hubble
expansion, whereas the second term is due to the peculiar velocity. 

In the case of \mbox{$v=v_{T}$}, for
\mbox{$t>t_{anh}$}, using (\ref{NE}), (\ref{ng}) and the usual value
of $\sigma_{T}$, (\ref{vt}) gives,

\begin{equation}
v_{T}=D\,\frac{t^{5/2}}{\xi^{3}}
\label{vT}
\end{equation}

where

\begin{equation}
D\sim 10^{-57}K^{2}GeV^{-3/2}
\label{D}
\end{equation}

Using (\ref{vT}), the evolution equation (\ref{dxi}) gives,

\begin{equation}
\xi(t)^{4} = (\frac{t}{t_{i}})^{2}\,\xi_{i}^{4}+
\frac{8}{3}\,v_{T}(t)\,\xi^{3}t\,[1-(\frac{t_{i}}{t})^{3/2}]
\label{xivT}
\end{equation}

The evolution of the correlation length of the magnetic field
configuration is described initially by the Alfven expansion equation
(\ref{xivA}) until the moment when \mbox{$\xi\sim l_{T}$}. From then
on the growth of $\xi$ continues according either to (\ref{xivA}) or
to (\ref{xivT}), depending on the relative magnitudes of the
velocities $v_{A}$ and $v_{T}$. Using the above, we can calculate the
scale $\xi_{eq}$ of the correlated domains at $t_{eq}$ and, thus,
calculate the rms magnetic field from (\ref{brms}).

\subsection{Diffusion}

An important issue, which should be considered, is the
diffusion length of the freezing of the field. Indeed, the assumption
that the field is frozen into the plasma corresponds to neglecting the
diffusion term of the magnetohydrodynamical induction equation
\cite{jack}, 

\begin{equation}
\frac{\partial {\bf B}}{\partial t}=\nabla\times({\bf v}\times{\bf B})
+ \sigma^{-1}\nabla^{2}{\bf B} \label{XB}
\end{equation}
where {\bf v} is the plasma velocity and $\sigma$ is the
conductivity. In the limit of infinite conductivity 
the diffusion term of (\ref{XB}) vanishes and the
field is frozen into the plasma on all scales. However, if
$\sigma$ is finite then spatial variations of the magnetic field
of lengthscale $l$ will decay in a diffusion time,
\mbox{$\tau\simeq\sigma l^{2}$} \cite{jack}. Thus, the field at a
given time $t$ can be considered frozen into the plasma only over the
diffusion scale, 

\begin{equation}
l_{d}\sim\sqrt{\frac{t}{\sigma}} \label{ld}
\end{equation}

If \mbox{$l_{d}>\xi$}, the magnetic field
configuration is expected, in less than a Hubble time, to become
smooth on scales smaller than $l_{d}(t)$. Thus, in this case, it is
more realistic to consider a field configuration with
coherence length $l_{d}$ and magnitude of the coherent magnetic
field $\overline{B_{cd}}$, where
\mbox{$\overline{B_{cd}}=B_{cd}/n_{d}$} is the flux-averaged initial
magnetic field over \mbox{$n_{d}\equiv l_{d}/\xi_{i}$} number of
domains. 

An estimate of the plasma conductivity is necessary
to determine the diffusion length. The current density in the
plasma is given by, \mbox{${\bf J}=ne{\bf v}$}, where $n$ is the
number density of the charged particles. The velocity {\bf v} acquired
by the particles due to the electic field {\bf E}, can be estimated as
\mbox{${\bf v}\simeq e{\bf E}\tau_{c}/m$}, where $m$ is the particle
mass and \mbox{$\tau_{c}=l_{m\!f\!p}/v$} is the timescale of collisions.
Since the mean free path of the particles is given by,
\mbox{$l_{m\!f\!p}\simeq 1/n\sigma_{c}$}, the current density is, 
\mbox{${\bf J}\simeq e^{2}{\bf E}/mv\sigma_{c}$}, where $\sigma_{c}$
is the collision cross-section of the plasma particles. Comparing with
Ohm's law gives for the conductivity \cite{jack}, \cite{plasma},

\begin{equation}
\sigma\simeq\frac{e^{2}}{mv\sigma_{c}}
\label{sigma}
\end{equation}

The collision cross-section is given by the Coulomb formula
\cite{plasma},

\begin{equation}
\sigma_{c}\simeq\frac{e^{4}}{T^{2}}\,\ln\Lambda
\label{coulob}
\end{equation}
where \mbox{$\ln\Lambda\simeq\ln(e^{-3}\sqrt{T^{3}/n})$} is the Coulomb
logarithm. Thus, the behaviour of the
conductivity depends crucially on the temperature.

For low temperatures, \mbox{$T<m_{e}\simeq
1\,M\!eV$} (i.e. after $t_{anh}$), 
the velocity of the electrons is, \mbox{$v\sim\sqrt{T/m_{e}}$}. Thus, from
(\ref{sigma}) and (\ref{coulob}) the conductivity is given by,

\begin{equation}
\sigma\sim\frac{1}{e^{2}}\sqrt{\frac{T^{3}}{m_{e}}}\,\frac{1}{\ln\Lambda}
\label{sL}
\end{equation}

For high temperatures, \mbox{$T\gg m_{e}$},
(\ref{ne}) suggests that, \mbox{$\ln\Lambda\sim
1$}. Also, the mass of the plasma particles is dominated by
thermal corrections, i.e. \mbox{$m\sim T$}, and \mbox{$v\sim 1$}.
Consequently, in this case, (\ref{sigma}) and (\ref{coulob}) give for the
conductivity,
\begin{equation}
\sigma\sim\frac{T}{e^{2}}
\label{s}
\end{equation}

Using the above results we can estimate the diffusion length. Indeed,
from (\ref{ld}), (\ref{sL}) and (\ref{s}) we obtain,

\begin{equation}
l_{d}\sim\left\{ \begin{array}{lr}
10^{8}GeV^{1/2}T^{-3/2} & \;\;\;\;\;T\geq 1\,M\!eV \\
 & \\
10^{8}GeV^{3/4}T^{-7/4} & \;\;\;\;\;T<1\,M\!eV
\end{array}\right.
\label{ldT}
\end{equation}

An important point to stress is that the diffusion
length is also increasing with time. {\em If \mbox{$l_{d}>\xi$} then the
size of the correlated domains is actually determined by the diffusion
length and it is the growth of the later that drives the evolution of
the magnetic field configuration}. 

At this point we could briefly discuss the behaviour of any {\em
electric} field, produced by the phase transition. As can be seen by
(\ref{gmn}) and (\ref{Btpsi}), the electric field, \mbox{$E^{\mu
}\equiv F^{\mu 0}$} is determined by the time derivative of the
Higgs--field VEV. Thus, strong currents are expected to arise at the
time of the transition, when the VEV of the Higgs--field falls rapidly
from zero to $\mu $. These are the currents that accompany the
creation of the magnetic field \cite{andrian}. However, after the
transition and during inflation, the VEV of the Higgs--field, at any
point in space, is more or less fixed and constant in time, since the
field configuration is comovingly frozen. Therefore, {\em there should
not be any significant electric field surviving the transition}. After
inflation this comoving picture begins to change but the magnetic
field decouples from the Higgs--field and, thus, any electric field
produced, by shifting of the magnetic field lines, is related to
plasma motion phenomena. Since such reorientations occur, we expect
small electric fields to be present in the form of electromagnetic
waves, which will diffuse and thermalize the gradient energy of the
magnetic field, that is reduced by its reorientation and alignment.

\section{At the end of inflation}

\subsection{The Reheating Temperature}
 The time $t_{end} $ when inflation ends could be determined by the
reheating temperature $T_{reh} $ with the use of the well known
relation, 
\begin{equation}
t_{end}\simeq 0.3\, g_{*}^{-1/2} (\frac{m_{pl} }{T_{reh}^{2} })
\label{tendT}
\end{equation}
where $g_{*} $ is the number of particle degrees of freedom, which, in
most models, is of order $10^{2} $ (e.g. in the standard model it is
106.75 whereas in the minimal supersymmetric standard model is 229).

The reheating temperature is usually estimated by \cite{Lidl},
\begin{equation}
\frac{T_{reh} }{m_{pl} } \approx 0.78\,\alpha^{1/4}
g_{*}^{-1/4}(\frac{H_{end} }{m_{pl} } )^{^{1/2} } \label{TrHe}
\end{equation}
where $H_{end} $ is the Hubble parameter at the end of inflation and
$\alpha $ is the reheating efficiency, which determines how much of
the inflaton's energy is going to be thermalised. 
Using (\ref{Hsqr}) and (\ref{fend}) we obtain,
\begin{equation}
H_{end} \simeq \frac{m}{\sqrt{3} }
\label{Hend}
\end{equation}
Subsituting to (\ref{tendT}) and (\ref{TrHe}) we find,
\begin{equation}
T_{reh}^{2} \approx 0.35\,\alpha^{1/2} g_{*}^{-1/2} m\,m_{pl} \label{Tral}
\end{equation}

In most inflationary models reheating is prompt, it is completed quickly and
$\alpha \approx 1$. In case of a quadratic inflaton potential,
however, as in false vacuum inflation, the reheating process could be
incomplete and extremely inefficient \cite{Kof4}.
However, the magnitude of the reheating inefficiency is still
an open question. Kofman {\em et al.} \cite{Kof3} suggest that the
reheating temperature would be of the order,
\mbox{$T_{reh} \sim 10^{-2} \sqrt{m\,m_{pl} }$} which, compared to
(\ref{Tral}), implies that \mbox{$\alpha \sim 10^{-4} $}. Shtanov, Trashen
and Brandenberger \cite{Bran} make a lower estimate, \mbox{$T_{reh}\sim m$}.

\subsection{Thermal fluctuations}

The Higgs--field, through the Higgs mechanism, provides the masses of
the particles after the GUT phase transition. Thus, it is in that way
coupled to the thermal bath of the particles. Therefore, at
reheating, this coupling  introduces thermal corrections to
the effective potential of the Higgs--field. Consequently, if the
reheating temperature is high enough, the configuration of the 
Higgs--field may be destroyed due to excessive thermal fluctuations.
This will eraze any magnetic field if the later has not been frozen
into the plasma already. The above will occur if the temperature
exceeds the well known Ginzburg temperature $T_{G} $. Moreover, if the
temperature exceeds a critical value $T_{c} $ there  is danger of
thermal restoration of the GUT symmetry itself.

The Ginzburg and the critical temperatures
are simply related \cite{Shel},
\begin{equation}
T_{c}-T_{G} \sim \lambda T_{c}
\end{equation}
Thus, for \mbox{$\lambda\leq 1$},
\begin{equation}
T_{G} \sim T_{c} \sim \sqrt{\lambda}\,\mu \label{TcTG}
\end{equation}

Therefore, it is very important to see if the temperatures during the
reheating process could exceed $T_{c} $. At this point it should be
noted that {\em the reheating temperature is not the highest
temperature achieved during the reheating process}. Indeed, as soon as
the field begins its coherent oscillations, the temperature
rises rapidly and assumes its maximum value \cite{TW}, \cite{KT},
\begin{equation}
T_{max} \simeq (V_{end}^{1/4} T_{reh} )^{1/2} \label{TmVe}
\end{equation}
where $V_{end} $ is the energy density of the inflaton at the end of
inflation. From (\ref{chao}) and (\ref{fend}) we obtain,
\begin{equation}
V_{end} \sim 0.1\,m^{2}m_{pl}^{2} \label{Vend}
\end{equation}
Thus,
\begin{equation}
T_{max} \sim (\sqrt{m\,m_{pl}}\;T_{reh})^{1/2} \label{Tmax}
\end{equation}
Therefore, in order to avoid symmetry restoration and any Ginzburg
phenomena we should have,
\begin{equation}
T_{max} < T_{c} \Rightarrow
T_{reh} < \frac{\lambda\mu^{2} }{\sqrt{m\,m_{pl} } } \label{muTm}
\end{equation}

If the reheating temperature exceeds the above value then the magnetic field
is thermally unstable and we are in danger of restoring the GUT symmetry.
However, if the field survives then its stability is ensured \cite{andrian}.

After reaching its highest value, $T_{max} $, the temperature slowly
decreases during the matter dominated era of the coherent inflaton
oscillations, until it falls to the value $T_{reh} $ when the universe
becomes radiation dominated.\footnote{We should mention that this
small period of matter domination is not taken into account in our
treatment due to the fact that its duration is very small compared to
the timescales considered and so we choose to ignore it for the sake
of simplicity.}

\section{Constraints}
\subsection{Constraints on the parameters}

If we assume that the observed density perturbations are due to
inflation, then we have from COBE \cite{modl},
\begin{equation}
\frac{\sqrt{8\pi } }{m_{pl} }\,m = 5.5\times 10^{-6}
\end{equation}
which yields,
\begin{equation}
m\sim 10^{13}GeV
\end{equation}

Other restrictions of the model imposed on
$\mu $, $\lambda $ and $\lambda' $ are \cite{modl},
\begin{equation}
0<\lambda, \lambda'\leq 1,\;\;\;\;\;\mu\leq \frac{m_{pl} }{\sqrt{8\pi } }
\sim 10^{18}\,GeV \label{llmu}
\end{equation}

An additional constraints for the $\lambda $ are established by the
inflaton domination condition (\ref{infdom}). Also
the ratio of the $\lambda$'s
can be determined by (\ref{finst}) and (\ref{No}) with the
reasonable assumption that the number of e-foldings of inflation after
the phase transition is of order \mbox{$N\sim 10$},
\begin{equation}
\frac{\lambda }{\lambda' } \sim (\frac{m_{pl} }{\mu })^{^{2} }
\label{lls}
\end{equation}

\subsection{The Galactic Dynamo constraint}

 From the present understanding of the galactic dynamo process \cite{zeld}
it follows that, in order for a primordial magnetic field to be the
seed for the currently observed galactic magnetic field, it should be
stronger than $10^{-19}\,Gauss $ at the time of galaxy formation, on a
comoving scale of a protogalaxy (\mbox{$\sim 100\,kpc$}). 

Since the gravitational collapse of the protogalaxies enchances their
frozen-in magnetic field by a factor of
\mbox{$(\rho_{G}/\rho_{c})^{2/3}\sim 10^{3}$} (where
\mbox{$\rho_{G}\sim 10^{-24}g\,cm^{-3}$} is the typical mass density
of a galaxy and \mbox{$\rho_{c}\simeq 2\times 10^{-29}\Omega
h^{2}g\,cm^{-3}$} is the current cosmic mass density), the above seed
field corresponds to an field of the order of \mbox{$\sim
10^{-22}Gauss$} over the comoving scale of \mbox{$\sim 1\,M\!pc$}.
With the assumption that the rms field scales as $a^{-2}$ with the
expansion of the universe (\mbox{$a\propto t^{2/3}$} for the matter
era), we find that the required magnitude of the seed field at
$t_{eq}$ is \mbox{$\sim 10^{-22}Gauss\times (t_{gc}/t_{eq})^{4/3}\sim
10^{-20}Gauss$}, where \mbox{$t_{gc}\sim10^{15}sec$} is the time of
the gravitational collapse of the galaxies. 

The above justify our choice to calculate the magnetic field at
$t_{eq} $ over the comoving scale of \mbox{$1\,M\!pc$} and consider
the constraint,

\begin{equation}
B^{eq}\geq 10^{-20}\,Gauss \label{Zbeq}
\end{equation}

 From recombination onwards the non-linear nature of structure
formation is very difficult to follow. Indeed, there exist a numerous
collection of different models. A strong primordial magnetic field
could influence in various ways some of these models, possibly with a
positive rather than a negative effect.

\subsection{The Nucleosynthesis constraint}

One upper bound to be placed on the magnetic field at $t_{eq} $ is
coming from nucleosynthesis. This has been
studied in detail by Cheng {\em et al.} \cite{nuc1}. They conclude
that, at \mbox{$t_{nuc}\sim 1\,sec$}, the magnetic field should not be
stronger than,

\begin{equation}
B^{nuc}\leq 10^{11}Gauss
\label{Bnuc}
\end{equation}
on a scale larger than \mbox{$\sim 10^{4}cm$}. A more recent
treatment by Kernan {\em et al.} \cite{nuc2} relaxes the bound by about
an order of magnitude, \mbox{$B^{nuc}\leq
e^{-1}(T_{\nu}^{nuc})^{2}\sim 10^{12}Gauss$}, where $T_{\nu}$ is the
neutrino temperature and $e$ is the electric charge. This bound is
valid over all scales. Similar results are also reached by Grasso and
Rubinstein \cite{grasso}. 

\subsection{Energy Density constraints}

Constraints are also induced by ensuring that the energy density of the
magnetic field is less than the energy density of the universe.
During the inflationary period, due to inflaton domination, the energy
density of the universe is mainly in the inflaton field. However, after
reheating and until $t_{eq} $, the energy density of the Universe is
just the radiation energy density.

Thus, for the inflationary period we should verify that,

\begin{equation}
\frac{\rho_{B} }{\rho_{inf} }\ll 1
\end{equation}
where \mbox{$\rho_{B}\equiv B^{2}_{cd}/8\pi  $} and
\mbox{$\rho_{inf}\equiv V(\phi ) $} are the energy densities of the
magnetic and inflaton fields respectively.

The highest value of the above ratio corresponds to the time of the
phase transition since the magnetic field is rapidly diluted during
inflation, whereas the inflaton's potential energy remains almost
unchanged. Using (\ref{B0}) and (\ref{chao}) we
find the first energy density constraint,

\begin{equation}
\sqrt{m\phi_{0}}>\xi_{F}^{-1}
\label{ratio}
\end{equation}

After reheating, the expansion of the
universe dilutes the energy density
$\rho_{B}$ of the magnetic field,  
inside a correlated domain more effectively than the radiation
density, which scales as $a^{-4}$.
Therefore, it is sufficient to ensure that
$\rho_{B}(t)$ is less than the energy density $\rho(t)$ of radiation
at the time $t_{i}$ of the formation of the magnetic field
configuration. That is,

\begin{equation}
\rho_{B}(t_{i})\leq\rho(t_{i})\Rightarrow
B_{cd}^{i}\leq\frac{\sqrt{3}}{2}\,\frac{m_{pl}}{t_{i}}
\label{enrg}
\end{equation}
which is the second energy density constraint.

\subsection{The non-Abelian constraint}

During the electroweak era, the freezing of the magnetic field into
the electroweak plasma is not at all trivial to assume. Indeed, before
the electroweak transition, since the electroweak symmetry group
\mbox{$SU(2)\times U(1)_{Y}$} is still unbroken, there are four
apparent ``magnetic'' fields, three of which are non-Abelian.

It would be more precise, then, to refer only
to the Abelian (Hypercharge) part of the magnetic field, which
satisfies the same magnetohydrodynamical equations as the Maxwell
field of electromagnetism. The non-Abelian part of the field may not
influence the motion of the plasma due to the existence of a
temperature dependent magnetic mass, \mbox{$m_{B}\approx 0.28\,g^{2}T$}
(see for example \cite{Laza} and \cite{NATO}), which could screen the
field over the relevant lengthscales. 

The condition for this screening to be effective can be obtained by
comparing the screening length \mbox{$r_{S}\sim m_{B}^{-1}$} of the
non-Abelian magnetic fields with the Larmor radius of the plasma
motion \mbox{$r_{L}\sim\frac{mv}{gB}$}, where
\mbox{$m\sim\sqrt{\alpha}\,T$} (\mbox{$\alpha=g^{2}/4\pi $}) is the
temperature induced physical mass of the plasma particles,
\mbox{$g\simeq 0.3$} is the gauge coupling (charge) and $v$ is the
plasma particle velocity. If we assume thermal velocity distribution,
i.e. \mbox{$mv^{2}\sim T$}, we find,
\begin{equation}
R\equiv\frac{r_{L} }{r_{S} }\sim 10^{-2}\,\frac{T^{2} }{B_{cd} }
\label{R}
\end{equation}

If \mbox{$R\geq 1$} then our restriction to the
Abelian (Hypercharge) part of the magnetic field is well justified. 
This restriction will not cause any
significant change to our results since, at the electroweak
transition, the hypercharge field projects onto the photon through
the Weinberg angle (\ref{wein}), \mbox{$\cos\theta_{W}\approx 0.88$}.
If, however, \mbox{$R<1$} then the non-Abelian fields do affect the
plasma motion, and should be taken into account.
Since \mbox{$T\propto a^{-1}$} and $B_{cd}$ falls at least as rapid as
$a^{-2}$, $R$ is, in general, an increasing function of time. 
Thus, the constraint has to be evaluated at reheating.

\subsection{The monopole constraints}

Unfortunately, the mechanism, which we use to generate the primordial
magnetic field, could also produce
stable magnetic monopoles. Since these monopoles should not dominate
the energy density of the universe, we require that the the fraction
$\Omega_{M}$ of the critical density, contributed by the monopoles, to
be less than unity, that is \cite{KT},
\begin{equation}
\Omega_{M}\,h^{2}\simeq 10^{24} (\frac{n_{M}
}{s})(\frac{M}{10^{16}GeV})\leq 1
\label{WM}
\end{equation}
where \mbox{$M=4\pi\mu g^{-1}\sim 10\,\mu $} is the monopole mass,
$n_{M}$ is the monopole number density, $s$ is the entropy density of
the universe and $h$ is the Hubble constant in units of \mbox{$100\,
km/sec/Mpc$}. The ratio $n_{M}/s$ is a constant\footnote{we can ignore
monopole annihilations (see \cite{KT})} and can be evaluated at the
end of inflation $t_{end}$. Taking \mbox{$n_{M}\sim\xi_{end}^{-3}$},
where $\xi_{end}$ is the correlation length at that time, we have, 
\begin{equation}
\frac{n_{M} }{s}\simeq\frac{10^{2} }{\zeta^{3} }\,(\frac{T_{reh}
}{m_{pl} })^{3}
\label{nMs}
\end{equation}
where \mbox{$\zeta\equiv\xi_{end}H_{end}$} gives the correlation length as
a fraction of the Hubble radius. From (\ref{WM}) and (\ref{nMs}) we
find the first monopole constraint \cite{modl},
\begin{equation}
\zeta^{3}\geq 10^{11}(\frac{T_{reh} }{10^{14}GeV})^{3}\,(\frac{\mu
}{10^{15}GeV}) 
\label{mon1}
\end{equation}

Apart from the above, mass density constraint, another constraint is
the well known ``Parker bound'' \cite{parker}, which considers the
effect of galactic magnetic fields onto the magnetic monopole motion.
The flux of the monopoles is \cite{KT},
\begin{equation}
\Phi_{M} =\frac{1}{4\pi }\,n_{M} v_{M}\sim 10^{10}(\frac{n_{M}
}{s})(\frac{v_{M} }{10^{-3} })\,cm^{-2}sr^{-1}sec^{-1}
\label{flux}
\end{equation}
where $v_{M}$ is the monopole velocity. The monopoles are accelerated
by the galactic magnetic field \mbox{$B_{g}\sim 10^{-6}Gauss$} to
velocity,
\begin{equation}
v_{M}\simeq (\frac{2h_{M}B_{g}l}{M})^{1/2}\sim
10^{-3}(\frac{10^{16}GeV}{M})^{1/2} 
\label{vM}
\end{equation}
where \mbox{$l\sim 1\,kpc$} is the coherence length of the magnetic
field and \mbox{$h_{M}\sim e^{-1} $} is the magnetic charge of the
monopole.

The magnetic field ejects the monopoles from the galaxy, while
providing them with kinetic energy \mbox{$E_{K}\simeq hB_{g}l\sim
10^{11}GeV$}. Demanding that the monopoles do not drain the field
energy in shorter times than the dynamo timescale, i.e. the galactic
rotation period \mbox{$\tau\sim 10^{8}yrs$}, we find the constraint,
\begin{equation}
\frac{B_{g}^{2}/2}{\Phi_{M}E_{K}d}\geq\tau\Rightarrow \Phi_{M}\leq
10^{-15}cm^{-2}sr^{-1}sec^{-1}
\label{comp}
\end{equation}
where \mbox{$d\simeq 30\,kpc$} is the size of the galactic magnetic
field region. Using (\ref{nMs}) and from (\ref{flux}) and (\ref{comp})
we find the second monopole constraint,
\begin{equation}
\zeta^{3}\geq 10^{12}(\frac{T_{reh}
}{10^{14}GeV})^{3}\,(\frac{10^{15}GeV}{\mu })^{1/2}
\label{mon2}
\end{equation}

\subsection{Additional constraints and considerations}

Finally, we have to make sure that at the time the magnetic field is
formed the correlation length given by (\ref{xF}) is still inside the
horizon, that is,
\begin{equation}
H_{F}^{-1}\geq\xi_{F} \label{Hoxi}
\end{equation}
where $H_{F} $ is the Hubble parameter at the time of the formation of
the magnetic field. With great accuracy \mbox{$H_{F}\simeq H_{0}$},
where $H_{0}$ is the Hubble parameter at the time of the phase
transition. $H_{0} $ can be easily computed using (\ref{finst}),
(\ref{fdot}), and (\ref{roll}) for the potential (\ref{chao}).

\section{Evaluating}

In order to be consistent with our assumptions (e.g. inflaton
domination) we will consider the phase transition to take place at the
latest at, \mbox{$N\simeq 1$}. We first choose a set of typical
model--parameters.

\subsection{Choosing the values of the parameters}

As already mentioned, the mass of the inflaton field $m$ is determined by
COBE,
\begin{equation}
m\sim 10^{13}GeV
\end{equation}

For the selfcoupling of the Higgs--field $\lambda$ we choose the usual value,
\begin{equation}
\lambda\sim 1
\end{equation}

Inserting the above values into the inflaton domination condition
(\ref{infdom}) we find that the maximum value of $\mu$ is,
\begin{equation}
\mu\sim 10^{15}GeV
\end{equation}

Finally, the coupling $\lambda'$ between the Higgs--field and the inflaton
can be determined with the use of (\ref{lls}),
\begin{equation}
\lambda'\sim 10^{-8}
\end{equation}

\subsection{For all $N$}

Now that the parameters of the model are chosen, the only parameter
still to be determined is the number $N$ of e-foldings of inflation, which
remain after the phase transition. We will treat this as a free parameter,
link it with the resulting magnetic field and then try, with the use of the
constraints, to establish its extreme values. In that way we will be able
to fully examine the corresponding behaviour of the field at $t_{eq}$.

We begin by extracting some direct, $N$-independent results from the,
previously chosen, values of the model--parameters. From (\ref{muTm}) we find
that the upper bound for the reheating temperature is estimated to be,
\begin{equation}
T_{reh}\sim 10^{14}GeV
\end{equation}
which is in agreement with the estimates of Kofman {\em et al} \cite{Kof3}
and higher than the estimates of Shtanov {\em et al} \cite{Bran}.

Using (\ref{tendT}) this gives the time
when inflation ends,
\begin{equation}
t_{end}\sim 10^{-35}sec \label{tend1}
\end{equation}

With the use of (\ref{finst}) and (\ref{brac}), the assumption (\ref{asum})
is easily verified,
\begin{equation}
\Lambda\sim 10^{-2}\ll 1
\end{equation}

Now, for the correlation length, from (\ref{corl}) we find,
\begin{equation}
\xi_{0}\sim 10^{-14}GeV^{-1}
\end{equation}

From (\ref{finst}), and (\ref{tf}) we obtain,
\begin{equation}
\tau_{F}\sim 10^{-16}GeV^{-1}\ll\xi_{0}
\end{equation}

Thus, from (\ref{xifl}),
\begin{equation}
\xi_{F}^{(1)}\sim 10^{-14}GeV^{-1}
\end{equation}

Using (\ref{xifm}) we also find,
\begin{equation}
\xi_{F}^{(2)}\sim 10^{-14}GeV^{-1}
\end{equation}

Therefore, from (\ref{xF}) and the above we have,
\begin{equation}
\xi_{F}\sim\xi_{0}\sim 10^{-14}GeV^{-1}\sim 10^{-28}cm\sim 10^{-47}pc
\label{cor1} 
\end{equation}

Now, from (\ref{DN}) we find that,
\begin{equation}
\Delta N_{F}\sim 10^{-3}\ll 1
\end{equation}
and, therefore, the phase transition is very rapid.

We can, now, check on the horizon constraint (\ref{Hoxi}). The value of
$H_{F} $ is found to be,
\begin{equation}
H_{F}\simeq H_{0}\sim 10^{13}GeV
\end{equation}
Comparing with (\ref{cor1}) we see that the constraint is satisfied.
By using (\ref{cor1}) into (\ref{ratio}) we find,
that the first energy density constraint is also satisfied.

The initial magnetic field is found from (\ref{B0}) and (\ref{cor1}) to be,
\begin{equation}
B_{0}\sim 10^{47}Gauss \label{B01}
\end{equation}

Let us now evaluate the $N$-dependent quantities.

The correlation length at the end of inflation is,
\begin{equation}
\xi_{end}=\frac{a_{end} }{a_{0} }\xi_{F}\sim 10^{-14}e^{N}GeV^{-1}
\label{xiend}
\end{equation}
where we have used (\ref{Nf1f2}), (\ref{No}) and (\ref{cor1}).

From the above and considering also the fact that, during inflation,
the magnetic field configuration is comovingly frozen,  we find that
the magnitude of the magnetic field inside a correlated domain is
given by, 

\begin{equation}
B_{cd}^{end}\sim 10^{47}e^{-2N}Gauss
\label{bcdi}
\end{equation}

Evaluating, (\ref{R}) at the
end of inflation, we find,
\begin{equation}
R\sim 0.1\,e^{2N}\geq 1\;\;\;\;\;\;\;\;\;\;\mbox{for $N\geq 1$}
\end{equation}
and the non-Abelian constraint is satisfied for all $N$.

Using (\ref{tend1}) and (\ref{bcdi}) we can show from
(\ref{enrg}) that the second energy density constraint is also
satisfied for all $N$.

From (\ref{K}) we find,
\begin{equation}
K\sim 10^{8}e^{-N}GeV^{1/2}
\label{K1}
\end{equation}
We evaluated the above also at the end of inflation, using (\ref{mgfd}),
(\ref{tend1}) and (\ref{xiend}).

At early times the correlated domains are opaque to radiation and,
thus their growth is determined by (\ref{xivA}) with
\mbox{$t_{i}\rightarrow t_{end}$}. The domains remain opaque at least until
the time of the electron pair annihilation 

At annihilation \mbox{$t_{anh}\sim 0.1\,sec$}, (\ref{xivA}) gives,

\begin{equation}
\xi_{anh}\sim\left\{ \begin{array}{lr}
10^{3}e^{N}GeV^{-1} & \;\;\;\;\;N> 15 \\
 & \\
10^{13}e^{-N/2}GeV^{-1} & \;\;\;\;\;1\leq N\leq 15 \end{array}\right.
\label{xianh}
\end{equation}

Comparing with the photon mean free path it is evident that, 
\mbox{$\xi_{anh} >l_{T}\sim 10^{5}GeV^{-1}$}, 
for all $N$.

However, after annihilation, $l_{T}$ increases drastically in size,
\mbox{$l_{T}(t_{anh})\sim 10^{15}GeV^{-1}$}. 
Comparing this value with (\ref{xianh}) we find that, at \mbox{$T\sim
1\, MeV$}, the correlated domains become transparent to radiation for
\mbox{$N<28$}. Thus, for \mbox{$N\geq 28$} the
Alfven expansion continues after pair annihilation, whereas, for
\mbox{$N<28$} the Thomson scattering effect has to be taken into
account.

\bigskip

CASE 1:\ {\em For $1\leq N<28$}

\nopagebreak[4]

\bigskip

\nopagebreak[4]

If \mbox{$N<28$} then, after $t_{anh}$ the Thomson effect has to be
taken into account. 

The Alfven velocity $v_{A}$ at $t_{anh}$ is found by (\ref{vA}) with
the use of (\ref{K1}),
\begin{equation}
v_{A}(t_{anh})\sim\left\{
\begin{array}{lr}
0.1\,e^{-2N} & \;\;\;\;\;N>15\\
 & \\
10^{-11}e^{-N/2} & \;\;\;\;\;1\leq N\leq 15
\end{array}\right.
\end{equation}

Similarly, the Thomson velocity $v_{T}$ at $t_{anh}$ is found by
(\ref{vT}) with the use of (\ref{K}) and (\ref{D}),
\begin{equation}
v_{T}(t_{anh})\sim\left\{
\begin{array}{lr}
10^{10}e^{-5N} & \;\;\;\;\;N>15\\
 & \\
10^{-20}e^{-N/2} & \;\;\;\;\;1\leq N\leq 15
\end{array}\right.
\end{equation}

From the above it is straightforward that, for
\mbox{$N<28$}, \mbox{$v_{T}(t_{anh})\leq v_{A}(t_{anh})$}. Therefore, 
after $t_{anh}$, the evolution of the correlated domains is determined
by the Thomson effect. If we assume that the Alfven expansion does not
take over again until $t_{eq}$ then the correlation length at that
time can be obtained by (\ref{xivT}),

\begin{equation}
\xi_{eq}\sim\left\{ 
\begin{array}{lr}
10^{9}e^{N}GeV^{-1} & \;\;\;\;\;19\leq N<28 \\
 & \\
10^{21}e^{-N/2}GeV^{-1} & \;\;\;\;\;1\leq N<19
\end{array}\right.
\label{xieq1}
\end{equation}

Using this value we can verify that the Thomson velocity remains
always smaller than the Alfven velocity until $t_{eq}$. 
Physically, (\ref{xieq1}) implies that, if \mbox{$N\geq 19$} the
damping of the growth of the correlated domains, after $t_{anh}$ is so
effective that the Hubble term dominates their evolution.
However, for \mbox{$1\leq N<19$}, the Thomson
velocity, although small, is still capable of outshining the Hubble
term. 

\bigskip

CASE 2:\ {\em For $N\geq 28$}

\nopagebreak[4]

\bigskip

\nopagebreak[4]

For high values of $N$ the magnetic field is so much diluted by
inflation that the Alfven or Thomson expansions are insignificant. The
growth of the correlated domains is driven solely by the Hubble expansion
and, thus,
\begin{equation}
\xi_{eq}\simeq\sqrt{\frac{t_{eq} }{t_{anh} }}\,\xi_{anh}\sim
10^{9}e^{N}GeV^{-1}\;\;\;\;\;\;\;\;\;\;N\geq 28
\label{xieq2}
\end{equation}

In total, (\ref{xieq1}) and (\ref{xieq2}) suggest 
the following behaviour for the correlation
length at $t_{eq}$,

\begin{equation}
\xi_{eq}\sim\left\{ 
\begin{array}{lr}
10^{9}e^{N}GeV^{-1} & \;\;\;\;\;N\geq 19 \\
 & \\
10^{21}e^{-N/2}GeV^{-1} & \;\;\;\;\;1\leq N<19
\end{array}\right.
\end{equation}

From (\ref{ldT}) we find that at $t_{eq}$ the diffusion
length is, \mbox{$l_{d}^{eq}\sim 10^{23}GeV^{-1}$}. Comparing with the
above we see that \mbox{$l_{d}^{eq}>\xi_{eq}$} for \mbox{$N\leq 32$}.
Thus, the dimensions of the correlated domains at $t_{eq}$ are
actually given by,

\begin{equation}
\xi_{eq}\sim\left\{ 
\begin{array}{lr}
10^{9}e^{N}GeV^{-1} & \;\;\;\;\;N>32 \\
 & \\
l_{d}^{eq}\sim 10^{23}GeV^{-1} & \;\;\;\;\;1\leq N\leq 32
\end{array}\right.
\label{xieq}
\end{equation}

\subsection{The magnetic field's range of values}

We are, now in the position to calculate the magnetic field strength
at $t_{eq}$. From (\ref{K}) and (\ref{xieq}) we have,

\begin{equation}
B_{cd}^{eq}\sim\left\{ 
\begin{array}{lr}
10\,e^{-2N}Gauss & \;\;\;\;\;N>32 \\
 & \\
10^{-13}e^{-N}Gauss & \;\;\;\;\;1\leq N\leq 32
\end{array}\right.
\label{Bcd}
\end{equation}

Also, from (\ref{n}) and (\ref{xieq}) we get,

\begin{equation}
n\sim\left\{ \begin{array}{lr}
10^{24}e^{-N} & \;\;\;\;\;N>32 \\
 & \\
10^{10} & \;\;\;\;\;1\leq N\leq 32
\end{array}\right.
\label{n1}
\end{equation}

With the use of the above, in view also of (\ref{brms}), we can
immediately find the rms value of the field at $t_{eq}$ for a given $N$,

\begin{equation}
B^{eq}_{rms}\sim\left\{
\begin{array}{lr}
10^{-11}e^{-3N/2}Gauss & \;\;\;\;\;N>32 \\
 & \\
10^{-18}e^{-N}Gauss & \;\;\;\;\;1\leq N\leq 32
\end{array}\right.
\label{beq1}
\end{equation}

As can be seen from (\ref{beq1}), the maximum rms value
of the field at $t_{eq}$ corresponds to \mbox{$N=1$}, 

\begin{equation}
(B^{eq}_{rms})_{max}\sim 10^{-18}Gauss
\end{equation}

For the minimum value of the field we just employ the galactic dynamo
constraint (\ref{Zbeq}). This gives,
\begin{equation}
N_{max}\simeq 5
\end{equation}

Thus, the range of values of the magnetic field is,
\begin{eqnarray}
5\; \geq & N & \geq \; 1 \nonumber \\
10^{-20}Gauss \leq & B^{eq} & \leq 10^{-18}Gauss
\end{eqnarray}

The above result, however, are still subject to the
nucleosynthesis and monopole constraints.

\subsection{Nucleosynthesis and monopole constraints on $N$}

Since nucleosynthesis occurs very near the electron pair annihilation
we will assume, for simplicity, that the correlation length
$\xi_{nuc}$ at \mbox{$t_{nuc}\sim 1\,sec$} is approximately equal to
the one at annihilation, i.e.
\mbox{$\xi_{nuc}\sim\xi_{anh}$}.\vline\footnote{The magnitude of
$\xi_{nuc}$ does not affect the results when the Hubble term in the
evolution equations is subdominant. In the opposite case our
assumption perturbs the results by an order of magnitude in the values
of $\xi_{eq}$ and $n$ but less than an order of magnitude in the value
of the rms magnetic field, since the later depends on
\mbox{$1/\sqrt{n}$} [see (\ref{Brms})].} 

The diffusion length at $t_{anh}$ is found by (\ref{ldT}) to be,
\mbox{$l_{d}^{anh}\sim 10^{13}GeV^{-1}$}. Thus, from (\ref{xianh}) we
have,

\begin{equation}
\xi_{nuc}\sim \xi_{anh}\sim\left\{ 
\begin{array}{lr}
10^{3}e^{N}GeV^{-1} & \;\;\;\;\;N>23 \\
 & \\
\l_{d}^{anh}\sim 10^{13}GeV^{-1} & \;\;\;\;\;1\leq N\leq 23 
\end{array}\right.
\label{xianh1}
\end{equation}

Inserting the above into (\ref{K}) and with the use of (\ref{K1}) we
find, 

\begin{equation}
B_{cd}^{nuc}\sim\left\{ 
\begin{array}{lr}
10^{13}e^{-2N}Gauss & \;\;\;\;\;N>23 \\
 & \\
10^{3}Gauss & \;\;\;\;\;1\leq N\leq 23 
\end{array}\right.
\label{bnuc}
\end{equation}

The maximum value of the magnetic field is, 
\mbox{$B_{cd}^{nuc}(N=1)\sim 10^{2}Gauss$}.
Comparing with (\ref{Bnuc}), we see that the maximum
value of the field is well below the nucleosynthesis constraint and,
therefore, the constraint is not violated for any value of $N$.

Let us consider the monopole constraints.
Given the values of the model parameters and the assumed reheating
temperature, both of the monopole constraints (\ref{mon1}) and
(\ref{mon2}) reduce to, \mbox{$\zeta\geq 10^{4}$}. Using (\ref{Hend})
and (\ref{xiend}) we find that,
\begin{equation}
\zeta=\xi_{end} H_{end}\sim 0.1\,e^{N}
\end{equation}
and the constraints are satisfied only if \mbox{$N\geq 11$}.
Thus, a magnetic field strong enough to seed the galactic dynamo,
would violate the monopole constraints.

One way to overcome the monopole problem is to consider GUT--models
which do not admit monopole solutions, such as ``flipped''
$SU(5)$, i.e. the semi--simple group, \mbox{$SU(5)\times U(1)$}.

\section{Conclusions}

We have analysed the creation and evolution of a primordial magnetic field
in false vacuum inflation. We have shown that, in GUT-theories that do not
produce monopoles, a sufficiently strong primordial
magnetic field can be generated, provided that the phase transition
takes place no earlier than 5 e-foldings before the end of inflation.
Although the magnetic field produced is strong enough
to seed the dynamo process in galaxies, it does not violate any
of the numerous constraints imposed (apart from the monopole
constraint if applicable). 

Our results are sensitive to the reheating efficiency. Indeed, if reheating is
efficient, then the time of the end of inflation is earlier and the resulting
field diluted by the expansion of the universe. More importantly, though, if
the reheating temperature is of the order of the critical temperature or the
Ginzburg temperature, then the magnetic field will be erazed. Fortunatelly,
this does not appear to be the case.

Finally, the strength of the magnetic field produced by our mechanism relies on
the value of $N$, i.e. on the moment that the phase transition occurs.
In turn, this depends on the exact values of the model--parameters.
Observational data on the primordial magnetic field could determine, or
constraint, these parameters. Experiments to detect such a field have
occasionally been suggested (see for example \cite{zoao} or \cite{natr}).
Not merely would the observation of a primordial field yield information on
false vacuum inflation, but it would also improve our understanding of the
galactic dynamo and of non-linear astrophysical processes in general.

\bigskip

\bigskip

{\noindent{\Large{\bf Acknowledgements}}}

\bigskip

This work was partly supported by PPARC, the Greek State Scholarship
Foundation (I.K.Y.) and the E.U. under the HCM program (CHRX-CT94-0423). 
We would like to thank \mbox{B. Carter},
\mbox{A.R. Liddle}, \mbox{D.N. Schramm}, \mbox{T. Vachaspati} and
especially \mbox{M.J. Rees} for discussions. Finally,
we thank CERN for the hospitality.

\end{document}